\documentstyle[aps,pra,floats,epsfig]{revtex}

\newcommand{\be}{\begin{equation}}
\newcommand{\ee}{\end{equation}}
\newcommand{\bea}{\begin{eqnarray}}
\newcommand{\eea}{\end{eqnarray}}
\newcommand{\bt}{\begin{tabular}}
\newcommand{\et}{\end{tabular}}
\newcommand{\ba}{\begin{array}}
\newcommand{\ea}{\end{array}}

\newcommand{\dy}{\displaystyle}
\newcommand{\rt}{\rightarrow}

\begin{document}




\title{ \hfill{{\small DSF$-$45/2001}} \\
        \hfill{{\small math-ph/0204040}} \\
        \hfill{{}} \\
Majorana transformation for differential equations}
\author{Salvatore Esposito}

\address{Dipartimento di Scienze Fisiche, Universit\`{a} di Napoli
``Federico II'' and Istituto Nazionale di Fisica Nucleare, Sezione di
Napoli
\\ Complesso Universitario di Monte S. Angelo,  Via Cinthia, I-80126
Napoli, Italy
\\ E-mail: Salvatore.Esposito@na.infn.it }

\maketitle

\begin{abstract}
We present a method for reducing the order of ordinary differential
equations satisfying a given scaling relation (Majorana scale-invariant
equations). We also develop a variant of this method, aimed to reduce the
degree of non-linearity of the lower-order equation. Some applications of
these methods are carried out and, in particular, we show that second-order
Emden-Fowler equations can be transformed into first-order Abel equations.
\\
The work presented here is a generalization of a method used by Majorana in
order to solve the Thomas-Fermi equation.
\end{abstract}

\vspace{1cm}


\vskip2pc
\section{Introduction}

\noindent
In a recent paper \cite{ME} we have described a method, originally due to
Majorana \cite{volumetti}, able to give the series solution of the
Thomas-Fermi equation (with appropriate boundary conditions) through only
one quadrature. Such a method, giving a (semi-analytic) parametric solution
of the considered equation, is based on a particular double change of
variables which transforms the second order Thomas-Fermi equation into a
first order equation, whose solution is then obtained by series expansion.
\\
Here we show that the transformation method, used by Majorana in that
particular case, applies to a large class of ordinary differential
equations as well, and prove a simple but general theorem for reducing the
order of these equations.
\\
The Majorana idea is a straigthforward generalization of known concepts and
to show this we briefly recall, in the following section, some definitions
and peculiarities of particular differential equations. In section III we
then introduce a new class of differential equations and give the method
for reducing the order of such equations. In section IV a variant of this
method is presented and in section V some applications are reported which
are particularly relevant in mathematical physics.

\section{Preliminaries}

\noindent
Let us consider a general differential equation of order $n$ in the
independent variable $x$ and dependent one $y$:
\begin{equation}\label{1}
  F \left( x,y,y^\prime,y^{\prime \prime}, \dots , y^{(n)} \right) \; = \; 0
~~~,
\end{equation}
where a prime $^\prime$ denotes differentiation with respect to $x$.
\\
Eq. (\ref{1}) is said to be an {\it autonomous equation} if the variable
$x$ does not appear explicitly:
\begin{equation}\label{2}
  F \left( x,y,y^\prime,y^{\prime \prime}, \dots , y^{(n)} \right) \; = \;
  F \left( y,y^\prime,y^{\prime \prime}, \dots , y^{(n)} \right) ~~~.
\end{equation}
In such a case, by changing the set of variables from $(x,y(x))$ to a novel
one $(y,u(y))$ through:
\begin{equation}
\ba{rcl}
\dy y^\prime &=& \dy u(y) \\ & & \\
\dy y^{\prime \prime} &=& \dy u(y) \, \frac{d u(y)}{dy} \\ & & \\
\dots & &
\ea
\end{equation}
where $u(y)$ is a given function of $y$, the considered differential
equation can always be reduced to an equation of order $n-1$ in the
independent variable $y$ and dependent one $u$ \cite{Ors}:
\begin{equation}
G \left( y,u,\frac{du}{dy},\frac{d^2u}{dy^2}, \dots ,
\frac{d^{n-1}u}{dy^{n-1}} \right) \; = \; 0 ~~~.
\end{equation}
A differential equation (\ref{1}) is, instead, {\it equidimensional-in-}$x$
if it is invariant under the transformation $x \rt \alpha x$ for any
$\alpha \neq 0$:
\begin{equation}\label{3}
  F \left( \alpha x,y,\alpha^{-1}y^\prime,\alpha^{-2}y^{\prime \prime}, \dots ,
  \alpha^{-n}y^{(n)} \right) \; = \;
  F \left( x,y,y^\prime,y^{\prime \prime}, \dots , y^{(n)} \right) ~~~.
\end{equation}
This equation can be transformed \cite{Ors} into an autonomous equation in
the variables $(z,y(z))$:
\begin{equation}
G \left( y,\frac{dy}{dz},\frac{d^2y}{dz^2}, \dots ,
\frac{d^{n}y}{dz^{n}} \right) \; = \; 0 ~~~,
\end{equation}
by changing the independent variable:
\begin{equation}
\ba{rcl}
\dy x &=& \dy e^z \\ & & \\
\dy x \frac{d}{dx} &=& \dy \frac{d}{dz} \\ & & \\
\dots & &
\ea
\end{equation}
Thus, equidimensional-in-$x$ equations of order $n$ can always be reduced
to differential equations of order $n-1$.
\\
{\it Scale-invariant equations} satisfy the property:
\begin{equation}\label{4}
  F \left( \alpha x,\alpha^c y,\alpha^{c-1}y^\prime,\alpha^{c-2}y^{\prime \prime}, \dots ,
  \alpha^{c-n}y^{(n)} \right) \; = \;
  F \left( x,y,y^\prime,y^{\prime \prime}, \dots , y^{(n)} \right) ~~~.
\end{equation}
(i.e. they are invariant for $x \rt \alpha x$, $y \rt \alpha^c y$) for any
value of $\alpha \neq 0$ and some value $c$, and they can be transformed
into equidimensional-in-$x$ equations \cite{Ors}:
\begin{equation}
G \left( x,u(x),\frac{du}{dx},\frac{d^2u}{dx^2}, \dots ,
\frac{d^{n}u}{dx^{n}} \right) \; = \; 0 ~~~,
\end{equation}
with $G$ obeying Eq. (\ref{3}), by performing the following change of the
dependent variable:
\begin{equation}\label{SI}
y(x) \; = \; x^c \, u(x) ~~~.
\end{equation}
Even in this case, scale-invariant equations of order $n$can be thus
reduced to equations of order $n-1$.
\\
Finally, differential equations which are invariant under the
transformation $y \rt \alpha y$ for any $\alpha \neq 0$ are said to be {\it
equidimensional-in-}$y$ or {\it homogeneous} equations:
\begin{equation}\label{5}
  F \left( x,\alpha y,\alpha y^\prime,\alpha y^{\prime \prime}, \dots ,
  \alpha y^{(n)} \right) \; = \;
  F \left( x,y,y^\prime,y^{\prime \prime}, \dots , y^{(n)} \right) ~~~.
\end{equation}
By changing  the dependent variable through:
\begin{equation}\label{EY}
y(x) \; = \; e^{u(x)} ~~~,
\end{equation}
it can be transformed into an equation of order $n-1$ in the variables
$(x,u(x))$ \cite{Ors}:
\begin{equation}
G \left( x,u(x),\frac{du}{dx},\frac{d^2u}{dx^2}, \dots ,
\frac{d^{n-1}u}{dx^{n-1}} \right) \; = \; 0 ~~~.
\end{equation}
Thus, we know that the order of a given differential equation can always be
reduced of one unit if this belongs to one of the four different classes
mentioned above.

\section{Majorana transformation}

\noindent
For future convenience we now consider scale-invariant equations from a
different point of view and introduce another class of differential
equations. Equation (\ref{1}) is said to be {\it Majorana scale-invariant}
if it is invariant for $x \rt \alpha^c x$, $y \rt \alpha y$ for any $\alpha
\neq 0$ and some value $c$:
\begin{equation}\label{6}
  F \left( \alpha^c x,\alpha y,\alpha^{1-c}y^\prime,\alpha^{1-2c}y^{\prime \prime}, \dots ,
  \alpha^{1-nc}y^{(n)} \right) \; = \;
  F \left( x,y,y^\prime,y^{\prime \prime}, \dots , y^{(n)} \right) ~~~.
\end{equation}
It is easy to prove the following proposition: {\it a Majorana
scale-invariant equation of order $n$ can always be reduced to a
differential equation of order $n-1$}. In fact, by changing the role of the
dependent and the independent variable $x \rt y$, $y \rt x$ Eq. (\ref{1})
can be transformed into:
\begin{equation}\label{7}
G \left( y, x(y),\frac{dx}{dy},\frac{d^2x}{dy^2}, \dots ,
\frac{d^{n}x}{dy^{n}} \right) \; = \; 0 ~~~.
\end{equation}
where now $G$ satisfies the condition (\ref{4}):
\begin{equation}
  G \left( \alpha y,\alpha^c x,\alpha^{c-1} \frac{dx}{dy},\alpha^{c-2}
  \frac{d^2x}{dy^2}, \dots , \alpha^{c-n} \frac{d^nx}{dy^n} \right) \; = \;
  G \left( y, x(y),\frac{dx}{dy},\frac{d^2x}{dy^2}, \dots ,
  \frac{d^{n}x}{dy^{n}} \right) ~~~,
\end{equation}
that is Eq. (\ref{7}) is scale-invariant and the order can be reduced of
one unit.
\\
The concept introduced above is not a really new one, since it is related
to that of scale-invariant equations. However, the reformulation of the
problem in these terms is useful for developing the method for the
implementation of order reduction, which is a generalization of that used
by Majorana in the framework of the Thomas-Fermi equation \cite{ME}.
\\
We now describe in detail such a method, which carries out the solution of
Eq. (\ref{1}), with $F$ satisfying Eq. (\ref{6}), as given in parametric
form:
\begin{equation}\label{8}
\left\{ \ba{rcl}
x &=& x(t) \\ & & \\ y &=& y(t)
\ea     \right.   ~~~.
\end{equation}
Let us assume that $x$ in Eq. (\ref{8}) depends on the parameter $t$
through the function $y(t)$ and, eventually, on $t$ itself:
\begin{equation}\label{9}
x \; = \; x(t,y) \; = \; x(t,y(t)) \; = \; x(t) ~~~.
\end{equation}
Since $(x(t),y(t))$ in Eq. (\ref{8}) is a solution of the considered
differential equation (\ref{1}), supposed to be Majorana scale-invariant,
Eqs. (\ref{8}) and (\ref{9}) must satisfy the relation (\ref{6}), meaning
that for any $\alpha \neq 0$ and a given value $c$ we have:
\begin{equation}\label{10}
x(t,\alpha y) \; = \; \alpha^c \, x(t,y) ~~~.
\end{equation}
This implies that $x(t,y)$ should be an homogeneous function of $y$:
\begin{equation}\label{11}
x(t,y) \; = \; x(t,1) \, y^c \; \equiv \; z \, y^c ~~~,
\end{equation}
where $z=z(t)$ can be considered as an arbitrary but given function of the
parameter $t$. Note that, in such a way, the only unknown function to be
determined in order to satisfy Eq. (\ref{1}) is $y(t)$, and the parametric
solution (\ref{8}) can be rewritten, after Eq. (\ref{11}), as:
\begin{equation}\label{12}
\left\{ \ba{rcl}
x &=& z(t) \, y^c(t) \\ & & \\ y &=& y(t)
\ea \right.   ~~~,
\end{equation}
with $z(t)$ an arbitrary but given function of $t$ and $c$ is determined by
Eq. (\ref{6}).
\\
We have now to translate the differential equation (\ref{1}) for $y(x)$
into an equation for $y(t)$ in (\ref{12}). In the following,
differentiation with respect to $t$ will be denoted with a dot $\dot{}$,
while a prime $^\prime$ refers to differentiation with respect to $x$ as
above. The $t$-derivatives of $x$ are, from Eq. (\ref{12}), as follows:
\begin{equation}\label{13}
\ba{rcl}
\dy \dot{x} &=& \dy \left( \dot{z} + c z \frac{\dot{y}}{y} \right) \,
y^c \; \equiv \; x_1 \left( t, y, \dot{y} \right) \\ & & \\
\dy \ddot{x} &=& \dy \left\{ \ddot{z} + 2 c \dot{z} \frac{\dot{y}}{y} + c z \left[
(c-1) \left( \frac{\dot{y}}{y} \right)^2 + \frac{\ddot{y}}{y} \right]
\right\} \; \equiv \; x_2 \left( t, y, \dot{y}, \ddot{y} \right) \\ & & \\
& & \dots \\ & & \\
\dy \stackrel{n {\cdot}}{x} &=& \dots \; = \; x_n \left( t, y, \dot{y}, \ddot{y}, \dots ,
\stackrel{n {\cdot}}{y} \right) ~~~.
\ea
\end{equation}
Using these expressions we can obtain the $x$-derivatives of $y$, which are
present in Eq. (\ref{1}), in terms of $t$, $y$ and its $t$-derivatives:
\begin{equation}\label{14}
\ba{rcl}
\dy y^{\prime} &=& \dy \frac{\dot{y}}{\dot{x}} \; \equiv \; y_1 \left( t, y, \dot{y}
\right) \\ & & \\
\dy y^{\prime \prime} &=& \dy \frac{\dot{x} \ddot{y} - \ddot{x} \dot{y}}{\dot{x}^3} \;
\equiv \; y_2 \left( t, y, \dot{y}, \ddot{y} \right) \\ & & \\
& & \dots \\ & & \\
\dy y^{(n)} &=& \dots \; = \; y_n \left( t, y, \dot{y}, \ddot{y}, \dots ,
\stackrel{n {\cdot}}{y} \right) ~~~.
\ea
\end{equation}
The differential equation for $y(t)$ is then obtained by substituting Eqs.
(\ref{12}), (\ref{13}) and (\ref{14}) into Eq. (\ref{1}):
\begin{equation}\label{15}
\hat{F} \left( t, y, \dot{y}, \ddot{y}, \dots , \stackrel{n {\cdot}}{y} \right) \; = \; 0
~~~,
\end{equation}
where:
\begin{equation}\label{16}
\hat{F} \left( t, y, \dot{y}, \ddot{y}, \dots , \stackrel{n {\cdot}}{y} \right) \; \equiv \;
F \left( x(t,y),y,y_1(t,y,\dot{y}), \dots , y_n( t, y, \dot{y}, \ddot{y},
\dots , \stackrel{n {\cdot}}{y}) \right) ~~~.
\end{equation}
Note that, from Eq. (\ref{12}), the function $x(t,y)$ is homogeneous (with
respect to $y$) of degree $c$, while the functions $y_k( t, y, \dot{y},
\ddot{y}, \dots , \stackrel{k {\cdot}}{y})$ are homogeneous of degree $1-kc$:
\begin{equation}\label{17}
y_k \left( t, \alpha y, \alpha \dot{y}, \alpha \ddot{y}, \dots , \alpha
\stackrel{k {\cdot}}{y} \right) \; = \; \alpha^{1-kc} \,
y_k \left( t, y, \dot{y}, \ddot{y}, \dots ,
\stackrel{k {\cdot}}{y} \right) ~~~,
\end{equation}
as required for the Majorana property (\ref{6}) to be satisfied. In
particular from this we also deduce that the differential equation in
(\ref{15}) is equidimensional-in-$y$, since:
\bea
&~&\hat{F} \left( t, \alpha y, \alpha \dot{y}, \alpha \ddot{y}, \dots ,
\alpha \stackrel{n {\cdot}}{y} \right) \; = \nonumber \\
&~& F \left( x(t,\alpha y),\alpha y,y_1(t,\alpha y,\alpha \dot{y}), \dots ,
y_n( t, \alpha y, \alpha \dot{y}, \alpha \ddot{y}, \dots ,
\alpha \stackrel{n {\cdot}}{y}) \right) \; = \nonumber \\
&~& F \left( \alpha^c x(t,y),\alpha y,\alpha^{1-c} y_1(t,y,\dot{y}),
\dots , \alpha^{1-nc} y_n( t, y, \dot{y}, \ddot{y}, \dots , \stackrel{n {\cdot}}{y})
\right) \; = \\
&~& F \left( x(t,y),y,y_1(t,y,\dot{y}), \dots , y_n( t, y, \dot{y}, \ddot{y},
\dots , \stackrel{n {\cdot}}{y}) \right) \; = \nonumber \\
&~& \hat{F} \left( t, y, \dot{y}, \ddot{y}, \dots , \stackrel{n {\cdot}}{y}
\right) ~~~,  \nonumber
\eea
that is the function $\hat{F}$ satisfies the relation (\ref{5}). We can
then use the transformation in (\ref{EY}) to reduce the order of the
equation. More precisely we set:
\begin{equation}\label{18}
y(t) \; = \; e^{\dy \int u(t) dt} ~~~,
\end{equation}
so that the $t$-derivatives of $y(t)$ are as follows:
\begin{equation}\label{19}
\ba{rcl}
\dy \dot{y} &=& \dy u \, y \; \equiv \; u_1(u) \, y \\ & & \\
\dy \ddot{y} &=& \dy \left( \dot{u} \, + \, u^2 \right) \, y \; \equiv \;
u_2(u,\dot{u}) \, y \\ & & \\ & & \dots \\ & & \\
\dy \stackrel{{n {\cdot}}}{y} &=& \dots \; \equiv \; u_n \left( u,\dot{u},\ddot{u},\dots
,\stackrel{(n-1) {\cdot}}{u} \right) \, y ~~~.
\ea
\end{equation}
The unknown function is now $u(t)$ and the differential equation of order
$n-1$, obeyed by this quantity, is obtained by substituting Eqs. (\ref{19})
into Eq. (\ref{15}):
\begin{equation}
  \hat{F} \left( t,y, u_1 y, u_2 y, \dots , u_n y \right) \; = \; 0
\end{equation}
or, by using the homogeneity of the function $\hat{F}$ ($\hat{F} (t,y, u_1
y, u_2 y, \dots , u_n y) = \hat{F}( t,1, u_1, u_2, \dots , u_n)$), in all
points where $y(t)$ is different from zero we have:
\begin{equation}\label{20}
\hat{F} \left( t,1, u_1, u_2, \dots , u_n \right) \; = \; 0 ~~~.
\end{equation}
In terms of the initial function $F$ in Eq. (\ref{1}), by noting that:
\begin{equation}\label{21}
\ba{rcl}
\dy x(t,1) &=& \dy z(t) \\ & & \\
\dy y_1(t,1,u_1) &=& \dy \frac{u_1}{x_1(t,1,u_1)} \; = \; \frac{u}{\dot{z} + c u z}
\; \equiv \; v_1(t,u) \\ & & \\
\dy y_2(t,1,u_1,u_2) &=& \dy \frac{x_1(t,1,u_1) u_2 - x_2(t,1,u_1,u_2) u_1}{x_1^3(t,1,u_1)}
\; = \\ & & \\ &=& \dy \frac{\dot{z} \dot{u} - \ddot{z} u + (1-2c) \dot{z} u^2 +
c(1-c) z u^3}{(\dot{z} + c u z)^3} \; \equiv \; v_2(t,u,\dot{u}) \\ & & \\
& & \dots \\
\dy y_n(t,1,u_1,u_2,\dots,u_n) &=& \dots \; \equiv \; v_n \left( t,u,\dot{u},\ddot{u},\dots,
\stackrel{(n-1) {\cdot}}{u} \right)
\ea
\end{equation}
and using Eq. (\ref{16}), we have the final equation for $u(t)$:
\begin{equation}\label{22}
F \left( z(t) , 1 , v_1(t,u), v_2(t,u,\dot{u}), \dots , v_n (
t,u,\dot{u},\ddot{u},\dots, \stackrel{(n-1) {\cdot}}{u}) \right) \; = \; 0 ~~~.
\end{equation}
Summarizing, {\it the parametric solution of a Majorana scale-invariant
differential equation of order $n$ has the form as in Eq. (\ref{12}) with
$z(t)$ an arbitrary but given function of the parameter $t$ and $y(t)$ is
written as in Eq. (\ref{18}), where $u(t)$ satisfies the differential
equation (\ref{22}) of order $n-1$ and the functions $v_k( t, u,
\dot{u}, \ddot{u},\dots, \stackrel{(k-1) {\cdot}}{u})$ are evaluated as in Eqs.
(\ref{21})}.

\section{The method of the auxiliary function}

\noindent
Depending also on the choice for the function $z(t)$, it could happen that
Eq. (\ref{22}), although is of order $n-1$, is too much hard to be solved.
In some cases the following procedure can be used to reduce the degree of
non-linearity of Eq. (\ref{22}).
\\
Let us perform a change of the dependent variable:
\begin{equation}\label{23}
u ~~~ \longrightarrow ~~~ v \; = \; \frac{u}{\dot{z} + c u z} ~~~.
\end{equation}
In this case we have:
\bea
   v_1 &=& v \nonumber
\\ v_2 &=& \left( 1 - c v z \right) \, \frac{\dot{v}}{\dot{z}} \, + \,
\left( 1 - c \right) \, v^2  \label{24} ~~~,
\\ & & \dots \nonumber
\eea
and the function $y(t)$ in Eq. (\ref{12}) is given now by:
\begin{equation}\label{25}
y(t) \; = \; e^{\dy \int \frac{v \dot{z}}{1 - c v z} d t} ~~~,
\end{equation}
where $v(t)$ satisfies the differential equation of order $n-1$:
\begin{equation}\label{26}
F \left( z, \, 1, \, v, \, (1-c v z) \frac{\dot{v}}{\dot{z}} + (1-c) v^2 ,
\, \dots \right) \; = \; 0 ~~~.
\end{equation}

\section{Applications}

\noindent
As an application of the Majorana method described above, let us consider
the Emden-Fowler equation:
\begin{equation}\label{27}
y^{\prime \prime} \; = \; x^a \, y^b
\end{equation}
(with $a,b$ two real numbers), which is of particular interest in
mathematical physics \cite{Bellman}. It satisfies the relation (\ref{6})
with:
\begin{equation}\label{28}
c \; = \; \frac{1-b}{a+2}  ~~~,
\end{equation}
so that the method may apply only for $a \neq -2$. In this case, Eq.
(\ref{22}) for the Emden-Fowler equation (\ref{27}) is:
\begin{equation}\label{29}
\frac{\dot{z} \dot{u} - \ddot{z} u + (1-2c) \dot{z} u^2 +
c(1-c) z u^3}{(\dot{z} + c u z)^3} \; = \; z^a
\end{equation}
with $c$ given in Eq. (\ref{28}). After some algebra we arrive at the
following first-order equation for $u(t)$:
\begin{equation}\label{30}
\frac{du}{dt} \; = \; \alpha(t) \, + \, \beta(t) \, u \, + \, \gamma(t) \, u^2 \, + \,
\delta(t) \, u^3
\end{equation}
where:
\begin{equation}\label{31}
\ba{rcl}
\dy \alpha(t) &=& \dy z^a \, \dot{z}^2 \\ & & \\
\dy \beta(t) &=& \dy 3c z^{a+1} \, \dot{z} \, + \, \frac{\ddot{z}}{z} \\ & & \\
\dy \gamma(t) &=& \dy 3 c^2 \, z^{a+2} \, + \, 2 c \, - \, 1 \\ & & \\
\dy \delta(t) &=& \dy c^3 \, \frac{z^{a+3}}{\dot{z}} \, + \, c(c-1) \, \frac{z}{\dot{z}}
~~~. \ea
\end{equation}
With the Majorana method we have thus transformed the Emden-Fowler equation
(\ref{27}) into an Abel equation (\ref{30}) of the first kind. Depending on
the particular problem to be solved, the final equation can be further
simplified with an appropriate choice for $z(t)$ which, however, cannot be
chosen equal to a constant (in this case $\dot{z} \neq 0$ and Eq.
(\ref{28}) would not be a differential equation for $u$).
\\
A relevant case is that of Emden-Fowler equation with $b=1$ for which, from
Eq. (\ref{28}), we have $c=0$ and Eq. (\ref{30}) reduces to a simpler
Riccati equation:
\begin{equation}
\frac{du}{dt} \; = \; z^a \, \dot{z}^2 \, + \, \frac{\ddot{z}}{z}  \, u
\, - \, u^2 ~~~,
\end{equation}
which for $z(t)=t$ becomes:
\begin{equation}
\frac{du}{dt} \; = \; t^a \, - \, u^2 ~~~.
\end{equation}
Another interesting particular case is that of Thomas-Fermi equation, which
is an Emden-Fowler equation with $a=-1/2$, $b=3/2$:
\begin{equation}
y^{\prime \prime} \; = \; \frac{y^{3/2}}{\sqrt{x}} ~~~.
\end{equation}
The corresponding first-order equation (\ref{30}) for $u(t)$, choosing:
\begin{equation}
z(t) \; = \; \left[ 12 (1-t) \right]^{2/3} ~~~,
\end{equation}
is:
\be \label{33}
\frac{du}{dt} \; = \;  \frac{16}{3(1-t)} \, + \, \left( 8 \, + \, \frac{1}{3(1-t)}
\right) \, u \, + \, \left( \frac{7}{3} \, - \, 4 t \right) \, u^2 \, - \, \frac{2}{3}
\, t (1-t) \, u^3   ~~~.
\ee
This equation was obtained by Majorana \cite{volumetti} in studying the
Thomas-Fermi equation.
\\
Emden-Fowler equations can also be analyzed by using the method of the
auxiliary function outlined in the previous section. In this case, Eq.
(\ref{26}) for $v(t)$ becomes:
\begin{equation}\label{34}
\frac{dv}{dt} \; = \; \frac{\dy \dot{z} \left[ z^a - (1-c) v^2 \right]}{1 - c v z}
~~~,
\end{equation}
where $c$ is given in Eq. (\ref{28}). Note that also in this case we cannot
choose $z(t) =$ constant ($\dot{z} \neq 0$) from Eq. (\ref{25}).
\\
Although Eqs. (\ref{34}) and (\ref{30}) are different, in the particular
case with $b=1$, and thus $c=0$, we again obtain a Riccati equation:
\begin{equation}
\frac{dv}{dt} \; = \; \dot{z} \, \left( z^a \, - \, v^2 \right) ~~~.
\end{equation}
Instead, following the method of the auxiliary function with :
\begin{equation}
z(t) \, = \; 12^{2/3} \, t^2 ~~~,
\end{equation}
the Thomas-Fermi equation can be transformed into the first-order equation:
\begin{equation}\label{35}
\frac{d \tilde{v}}{dt} \; = \; 8 \, \frac{t \tilde{v}^2 - 1}{1 - t^2 \tilde{v}}
\end{equation}
where, for simplicity, we have set $\tilde{v} = - 4 {\cdot} 12^{-1/3} \, v$.
Equation (\ref{35}) has been solved using series expansion by Majorana
\cite{volumetti}, and this leads to a semi-analytic general solution for
the Thomas-Fermi equation (for details see \cite{ME}).

\section{Conclusions and outlook}

\noindent
In this paper we have generalized a result, derived by Majorana \cite{ME},
\cite{volumetti} for solving the Thomas-Fermi equation, to a wide class of
(ordinary) differential equations, that of Majorana scale-invariant
equations, as defined in Sect. III. We have shown that the search for the
parametric solution of such equations of order $n$ can be restricted to
that for the solution of a differential equation of order $n-1$ (and to the
computation of one integral involving this solution). However, the main
result of this paper is not this proposition, which is a direct consequence
of known results, but, rather, the method and the transformations used to
obtain the lower order equation. This has been outlined in Sect. III and
the equation considered has been formally written in Eq. (\ref{22}). In
some case this differential equation could be highly non-linear, so that in
Sect. IV we have developed a variant of the method mentioned above,
employing a further transformation for the dependent variable involved.
Some other simplifications, depending on the particular problem considered,
can be achieved with a suitable choice for the arbitrary function $z(t)$
present in the parametric solution for the differential equation.
\\
As an illustration, both methods have been applied to reduce the order of
Emden-Fowler equations in Sect. V and, as a particular case, Thomas-Fermi
equation has been considered as well. Remarkably, by using the method of
Sect. III, we have shown that all second-order Emden-Fowler equations can
be transformed into first-order Abel equations of the first kind. Instead,
by using the method of the auxiliary function reported in Sect. IV, the
Thomas-Fermi equation can be transformed into a suitable first-order
equation which can be solved by series expansion.
\\
We believe that the transformation methods presented here deserve further
attention in view of their potential applications to scale-invariant
differential equations which are of interest for mathematical physics.

\acknowledgements
This paper takes its origin from the study of some handwritten notes by E.
Majorana, deposited at Domus Galileana in Pisa, and from enlightening
discussions with Prof. E. Recami and Dr. E. Majorana jr. My deep gratitude
to them as well as special thanks to Dr. C. Segnini of the Domus Galileana
are here expressed.

\end{document}